\documentclass{mem}
\usepackage{natbib}\usepackage{txfonts}\usepackage{balance}
\usepackage{graphicx}
\usepackage[a4paper]{hyperref}
\idline{75}{282}
\begin{document}
\def\teff{$T\rm_{eff }$}
\def\kms{$\mathrm {km s}^{-1}$}

\title{
Distant early-type galaxies: tracers of the galaxy mass assembly evolution
}

\subtitle{}

\author{
A. \,Cimatti\inst{1} 
}

  \offprints{A. Cimatti}
 
\institute{
Istituto Nazionale di Astrofisica --
Osservatorio Astrofisico di Arcetri, Largo E. Fermi 5,
I-50125 Firenze, Italy
\email{cimatti@arcetri.astro.it}
}

\authorrunning{Cimatti}

\titlerunning{Early-type galaxy evolution}

\abstract{
We review the most recent observational results on the formation
and evolution of early-type galaxies and their mass assembly 
by focusing on: the existence, properties and role of distant old, 
massive, passive systems to $z\sim2$, the stellar mass function 
evolution, the ``downsizing'' scenario, and the high-$z$ precursors 
of massive early-type galaxies.
}
\maketitle{}

\section{Introduction}

Early-type galaxies (ETGs) play a crucial role in cosmology.
They represent the most massive galaxies in the local Universe,
contain most of the stellar mass (${\cal M}$) and 
are the primary probes to investigate the cosmic history of galaxy mass
assembly. As ETGs are the most clustered galaxies, they are also 
fundamental in tracing the evolution of the large scale structures.
Because of the correlation between the black hole and galaxy bulge 
masses, ETGs are also important to investigate the co--evolution 
of normal and active galaxies.

Although ETGs at $z\sim0$ are rather ``simple'' and homogeneous systems in 
terms of morphology, colors, stellar population content and scaling 
relations \cite{alvio}, their formation and evolution is still a debated 
question. In the so called "monolithic'' collapse scenario \cite{egg,
tinsley}, ETGs converted most of the gas mass into stars at 
high redshifts (e.g. $z>3$) through an intense burst of star formation 
followed by passive evolution of the stellar population. However, in 
the modern scenario of CDM cosmology, the evolution of the gravitational 
perturbations implies that galaxies assembled their mass more gradually 
through hierarchical merging of CDM halos and make the monolithic collapse 
unlikely or even unphysical. For instance, the most recent $\Lambda$CDM 
hierarchical models predict that most of the ETG stellar mass 
is assembled at $0<z<1$ \cite{delucia}. The modern generation of galaxy
surveys can test the different predictions on the evolution of the number 
density, colors, merger types and rates, luminosity and mass functions,
and provide a crucial feedback to improve the theoretical simulations.

Despite the intense observational activity on ETG evolution, a general
consensus has not been reached yet. For instance, recent results based 
on optically-selected, moderately deep ($R<24$) samples of ``red sequence''
ETGs (e.g. COMBO-17, DEEP2) \cite{bell04,faber} suggest a strong 
evolution of the number density with $\phi^*(z)$ displaying an increase 
by a factor of $\sim$6 from $z\sim1$ to $z\sim0$, and a corresponding 
increase by a factor of 2 of their stellar mass density since $z\sim1$. 
Dissipationless ETG--ETG merging (also called "dry" merging) is advocated 
to explain how ETGs assembled at $0<z<1$ \cite{bell05,vd05}. 
HST optical, moderately deep ($I<24$) samples of morphologically--selected 
ETGs also suggest a substantial decrease of the number density with $n 
\propto (1+z)^{-2.5}$ \cite{ferre}, and spatially resolved color maps 
show that $\sim$30-40\% of ETGs at $0<z<1.2$ display variations in their 
internal color properties (e.g. blue cores and inverse color gradients) 
suggestive of recent star formation activity in a fraction of the ETG 
population \cite{men}. However, the optically-selected ($I<24$) VIMOS 
VLT Deep Survey (VVDS) does not confirm the above findings, and shows 
that the rest-frame $B$-band luminosity function of ETGs (selected based 
on the SED) is consistent with passive evolution up to $z\sim1.1$, 
while the number of bright ETGs seems to decrease only by $\sim$40\% from 
$z\sim0.3$ to $z\sim1.1$ \cite{zucca}. The overall picture is made more
controversial by recent results indicating a weak evolution of the 
stellar mass function \cite{fon04,caputi,bundy} and that ``dry mergers'' 
cannot represent the major contributors to the assembly of ETGs at $0<z<1$ 
\cite{bundy,masjedi}.

\section{The problem of selection effects}

After the discovery of the strong clustering of ETGs at $z\sim1$
\cite{da00}, it became clear that a major source of uncertainty 
was the ``cosmic variance'' \cite{da00,som04a}, and that its
effects could explain the discrepant results often found by 
narrow-field surveys, 
such as the different content of luminous high-$z$ ETGs in the HDF-North 
and HDF-South \cite{zepf,benitez}. 

Another important source of uncertainty is played by the heterogeneous 
criteria adopted to select and classify ETGs: deep vs. shallow surveys, 
optical vs. near-infrared, color vs. morphological vs. spectrophotometric
or spectral energy distribution (SED) selections. 
ETGs have strong k--corrections in the optical as they become rapidly 
very faint for increasing redshifts and can be easily missed in 
optically--selected samples (e.g. Maoz 1997), while the k--corrections 
are much smaller in the near-infrared. A possible example is given by recent 
results \cite{yamada} based on a large sample of ETGs at $z\sim1$ extracted 
from a $\sim 1$ square degree field in the Subaru/XMM-Newton Deep Survey 
. Despite the field as large as that covered in COMBO-17, the number 
density of ETGs at $z\sim1$ turns out to be substantially larger than 
that derived at $z=0.95$ in COMBO-17, probably due to the deeper and 
redder--band selection ($z^{\prime}<25$ vs. $R<24$).

Also the ``progenitor bias'' \cite{vd01} plays a role in 
the comparison between low and high redshift ETG samples. As the
the morphology of galaxies are expected to evolve, some low-$z$ 
ETGs were spiral galaxies or mergers at higher redshift. These 
young ETGs are included in low redshift samples, but drop out 
of the samples at high redshift. Therefore, the high redshift samples
are a biased subset of the low redshift samples, containing only 
the oldest progenitors of low-$z$ ETGs, i.e. the ones with 
morphological and spectral properties similar to the nearby ETGs. 

\section{High redshift ETGs}

What is the highest redshift at which we still find ETGs similar to
those observed at $z\sim0$ ? After the isolated case of a spectroscopically 
identified old ETG counterpart of a radio source at $z=1.55$ \cite{dunlop}, 
studies based on the HDF-South and HDF-North suggested the 
existence of a substantial number of photometric/morphological ETG
candidates up to $z\sim2$ \cite{treu98,benitez,stanford}. It was
with the recent spectroscopic surveys of near-infrared galaxy samples
like the K20 \cite{cim02b,mig}, GDDS \cite{abra}, 
TESIS \cite{sar03}, GRAPES \cite{da05a}, LCIRS \cite{doherty},
that high-$z$ ETGs were unveiled up to $z\sim2$. 

The distant ETGs spectroscopically identified at $1<z<2$ 
are very red ($R-K_s>5$, $I-H>3$ in Vega photometric system), 
show the spectral features of passively evolving old stars with ages of 
1-4 Gyr (Fig.1), and have large stellar masses with ${\cal M}>10^{11} 
M_{\odot}$ \cite{cim02a,cim03,cim04,mcc,sar03,sar05,da05a,yan04,fu}. 
Their spectral properties ETGs suggest a weak 
evolution in the 
range of $0<z<1$ \cite{mig} (Fig.2). The deepest optical images of 
ETGs at $0.5<z<1.3$ in the Hubble Ultra Deep Field coupled with the 
GRAPES ACS slitless spectra show that their stellar populations are 
rather homogeneous in age and metallicity and formed at redshifts 
$z_f\sim2-5$, and that the isophotal structure of $z\sim1$ ETGs obeys 
the correlations already observed among nearby elliptical galaxies, 
implying no significant structural differences between $z\sim0$ and 
$z\sim1$ \cite{pasquali}. Distant ETGs are also strongly 
clustered, with a comoving $r_0 \sim 10$ Mpc at $z\sim1$ measured for 
the Extremely Red Object (ERO) population \cite{daddi02,firth,roche,
brown,georga}, and similar to that of present-day luminous ETGs. 

\begin{figure}[]
\resizebox{\hsize}{!}{\includegraphics[clip=true]{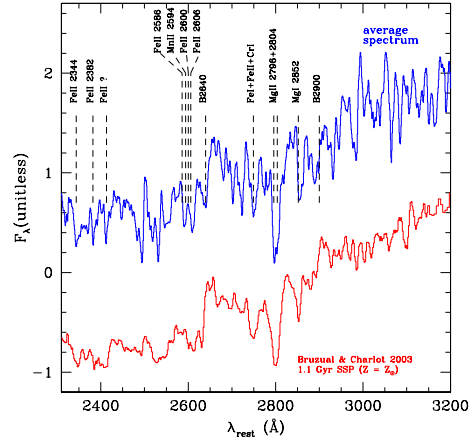}}
\caption{
\footnotesize
The observed average spectrum (top) of four massive ETGs 
at $1.6<z<2$ compared with a synthetic spectrum of a simple 
stellar population (bottom) \cite{cim04}.
}
\label{highz_etg_spec}
\end{figure}

The stellar masses of these galaxies are estimated by fitting their 
multi--band photometric SEDs with stellar population synthesis models 
\cite{fon04,sar05,long05,bundy}. Such ``photometric'' stellar masses 
are in reasonable agreement with the dynamical masses estimated from 
the absorption line velocity dispersion \cite{ssa,rettura}. ETGs dominate 
the high-luminosity and high-mass tails of the total luminosity and 
stellar mass functions up to $z\sim1$ \cite{pozz03,fon04,glaz04,bundy}.

\begin{figure}[]
\resizebox{\hsize}{!}{\includegraphics[clip=true,angle=-90]{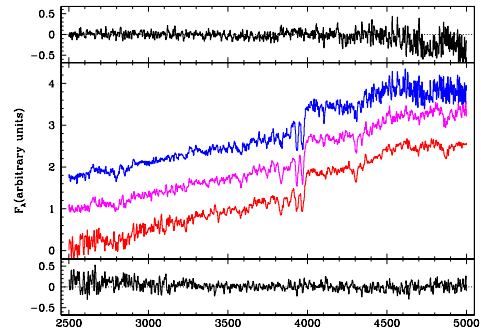}}
\caption{\footnotesize
Composite spectra of the K20 survey ETGs divided in three redshift bins 
(arbitrary flux scale) \cite{mig}. From bottom to top: $0<z<0.6$, $0.6<z<0.75$,
$0.75<z<1.25$. The upper and lower panels also show the difference spectra 
between the intermediate-redshift composite and the high-$z$ and low-$z$
composite spectra respectively.
}
\label{k20_etg_avg}
\end{figure}

However, due to the strong cosmic variance, the number density of high-$z$
ETGs is still so poorly constrained that the current data at $1<z<2$ are
consistent with a range from 10\% to 100\% of the local density of 
luminous ETGs \cite{cim04,da05a,sar05}. Old, passive, massive ETGs may 
exist even at $z>2$ \cite{labbe}, but their fluxes are typically beyond 
the current spectroscopic limits.

\section{Downsizing}

Distant ETGs have a higher stellar mass-to-optical light ratio 
(${\cal M}/L_R$) than late-type star-forming galaxies at the same 
redshift \cite{fon04} (Fig.3).  Moreover, the ${\cal M}/L_R$ 
shows an overall trend decreasing from low to high redshifts, 
but with a substantial scatter that can be interpreted as due to a 
wide range of formation redshifts and/or star formation histories
(see also McCarthy et al. 2001; Cimatti et al. 2003).
The typical ${\cal M}/L_R$ of more 
luminous objects ($M_R<-22$) is significantly larger than that of the fainter 
ones ($M_R>-22$) at low and intermediate redshifts (Fig.3). This implies 
that, while the ${\cal M}/L_R$ of the luminous population is consistent 
with either very short star-formation time-scales ($\tau$) or high 
formation redshifts ($z_{f} \geq 3$) (and some objects appear to require 
both), the less luminous population experienced a more recent history 
of assembly, as indicated by the larger $\tau$ and lower  $z_{f}$ required 
to reproduce the typical ${\cal M}/L_R$ (Fig. 3). Thus, the most luminous 
and massive ETGs reach the completion of star formation first, 
while less massive ones have a more prolonged star formation activity 
till later times. This {\it downsizing} scenario was first noted by 
\cite{cow96}, and it is now supported by several results which suggest 
that galaxy evolution is strongly driven by the galaxy mass. 
It is interesting to note that recent studies of ETGs at $z\sim0$
have reached consistent conclusions \cite{thomas}. 

\begin{figure}[]
\resizebox{\hsize}{!}{\includegraphics[clip=true]{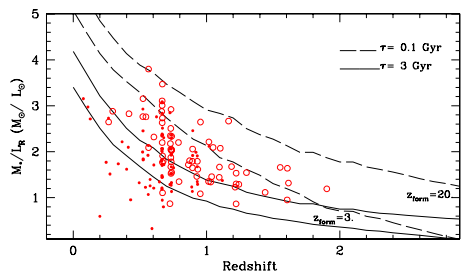}}
\caption{
\footnotesize
$M_{stars}/L_R$ ratio as a function of redshift for the spectroscopically 
identified ETGs in the the K20 sample. Large hollow and small filled 
circles indicate $M_R<-22$ and $M_R>-22$ respectively. Lines show 
$M_{stars}/L_R$ values predicted for a set of exponentially--decaying star 
formation rate models with $z_{f}=3$ (thin lines) or $z_{f}=20$ (thick 
lines) and time-scales of  $\tau = 0.1$ Gyr (dashed lines) and $\tau = 
3$ Gyr (solid lines) (Salpeter IMF and no dust extinction) \cite{fon04}.
}
\label{k20_mlratio}
\end{figure}

Additional clues on downsizing come from the most recent results
on the evolution of the Fundamental Plane (FP) both in the field and in
distant clusters \cite{van,treu,ssa,jorg}. The FP 
at $z\sim1$ shows already a remarkable small scatter and, with respect 
to the local FP, an offset and a possible rotation. The evolution of 
the overall 
FP can be represented by a mean change in the effective mass-to-light ratio, 
but with a rate depending
on the dynamical mass, being slower for larger masses. This differential 
evolution is consistent with stellar populations that formed at 
$z_f>2$ and $z_f \sim 1$ for high- and low-mass ETGs respectively.
Although, the relation between the measured $M/L$ evolution and mass is 
partially due to selection effects because the galaxies are selected by 
luminosity, not mass, the $M/L$ -- mass relation still holds even if this 
selection effect is taken into account. These results indicate that 
the fraction of stellar mass formed at recent times ranges from $<$1\% for 
$M>10^{11.5} M_{\odot}$ to 20\%-40\% below $M\sim 10^{11} M_{\odot}$, and
that there is no significant difference between the evolution of massive 
field and cluster ETGs (Fig.4). 

\begin{figure}[]
\resizebox{\hsize}{!}{\includegraphics[clip=true]{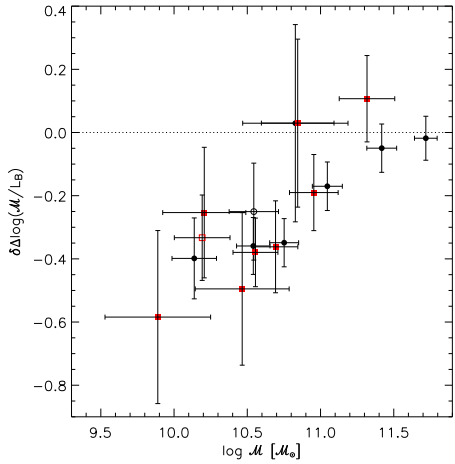}}
\caption{
\footnotesize
The differential evolution of the ${\cal M}/L_B$ ratio
with respect to that of old cluster massive galaxies at the same
redshift for ETGs at $z\sim1$ from the K20 survey \cite{ssa}.
}
\label{fp2}
\end{figure}

Recently it has been found that while the bright end of the cluster 
colour-magnitude relation is already built at $z\sim 0.8$, the faint 
end is still in the process of build-up \cite{tanaka} (see also De Lucia
et al. 2004). In contrast to this, the bright end of the field
colour-magnitude relation has been built all the way down to
the present-day, but the build-up at the faint end has not started yet. 
This suggests again the downsizing evolutionary pattern: massive 
galaxies complete their star formation first and the truncation of 
star formation is propagated to smaller objects as time progresses. 
This trend is likely to depend on environment since the build-up of 
the colour-magnitude relation is delayed in lower density environments. 
The evolution of galaxies took place earliest in massive galaxies and in
high-density regions, and it is delayed in less massive galaxies and in
lower density regions. 

Other evidences of the downsizing scenario come from the evolution of the 
stellar mass function (Fig.5-6). Several surveys consistently indicate that the 
high-mass tail of the stellar mass function (which is populated mostly by ETGs; 
Fig.6) evolves weakly from $z\sim 0.8-1.0$ to $z\sim0$ compared to the 
local stellar mass function and slower than the low-mass tail \cite{fon04,
dro05,caputi,bundy,pannella,franceschini}. 

\begin{figure}[]
\resizebox{\hsize}{!}{\includegraphics[clip=true,angle=-90]{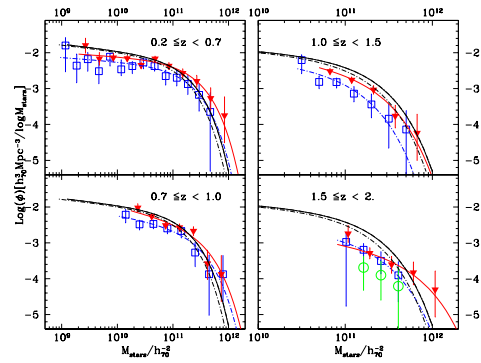}}
\caption{\footnotesize
Stellar mass functions in the K20 sample. Different symbols 
correspond to different methods adopted to estimate the stellar mass
\cite{fon04}. In the highest redshift bin, large hollow circles correspond 
to objects with only spectroscopic redshift. The thick solid line is 
the local galaxy mass function \cite{cole2001}. 
}
\label{k20_mf1}
\end{figure}

\begin{figure}[]
\resizebox{\hsize}{!}{\includegraphics[clip=true]{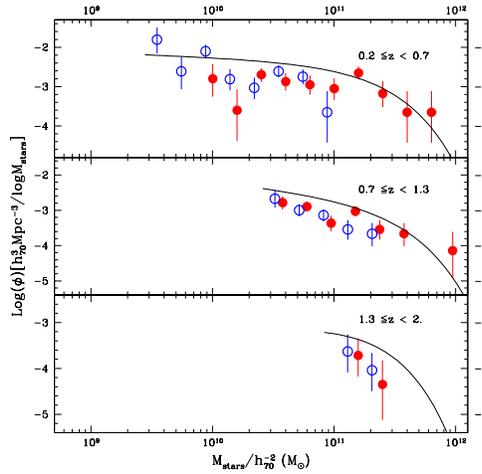}}
\caption{\footnotesize
Stellar mass functions in the K20 sample for
different spectral types. Empty points correspond to late
spectral types, filled to ETG spectral types.  The solid lines
show the Schechter fits to the total mass function of the K20
sample at the corresponding redshifts \cite{fon04}.
}
\label{k20_mf2}
\end{figure}

Recently, Bundy et al. (2006) confirmed this downsizing evolution of 
the stellar mass function per galaxy type up to $z\sim1$ by using 
the DEEP2 spectroscopic sample. In addition,
they identified a mass limit, ${\cal M}_Q$, above which star formation 
appears to be ``quenched''. This threshold mass evolves as $\propto
(1+z)^{4.5}$, i.e. it decreases with time by a factor of ~5 across the
redshift range of $0.4<z<1.4$, while no significant correlation has been 
found between downsizing and environment, with the exception of the most 
extreme environments (i.e. the ones hosting higher numbers of ETGs), where
the downsizing seems to be accelerated. Also the spectral properties of
ETGs support a downsizing scenario: Fig.7 shows that, at fixed redshift, 
the D4000 continuum break of ETGs (taken as an age indicator of the 
stellar population) is stronger in more luminous systems, suggesting
a higher formation redshift of the most massive ETGs \cite{mig}.

\begin{figure}[]
\resizebox{\hsize}{!}{\includegraphics[clip=true]{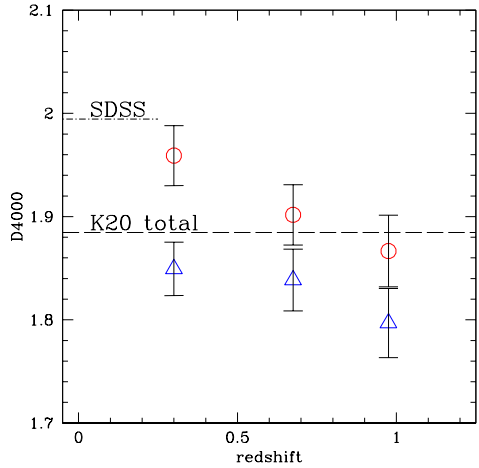}}
\caption{
\footnotesize
The 4000~\AA~ break (D4000) in K20 ETG composite spectra \cite{mig}.
The ETGs in each redshift bin have been divided in two equally populated 
groups, according to their luminosity. Circles and triangles indicate the 
brighter and fainter subsamples respectively. The dashed line indicates 
the D4000 value measured in the average spectrum of all ETGs, whereas 
the dashed-dotted line is the D4000 measured in the SDSS composite of 
ETGs \cite{eisen}. 
}
\label{d4000}
\end{figure}

Additional evidence supporting the downsizing scenario comes
from the results on the star formation history properties.
Feulner et al. (2005) found that, at all redshifts, lower mass galaxies show
higher specific star formation rates (SSFR = $SFR/{\cal M}$)
than higher mass galaxies, and that the highest mass galaxies contain
the oldest stellar populations at all redshifts. With increasing
redshift, the SSFR for massive galaxies increases by a factor of ~10,
reaching the era of their formation at $z\sim2$ and beyond. These findings
can be interpreted as evidence for an early epoch of star formation in
the most massive galaxies and for ongoing star formation activity in
lower mass galaxies. Similar results have been found also in other
recent works \cite{jun,perez}. 

\section{Comparison with model predictions}

These high-$z$ old and massive ETGs are very rare objects in most 
N-body + semi-analytic simulations published to date. The predicted 
number densities are much lower than the observed ones (up to a
factor of 10 in some cases), with this discrepancy becoming larger for 
increasing redshift \cite{som04b}. However, as there are enough dark 
matter halos to host these unexpected massive galaxies 
\cite{fon04}, what becomes crucial is to understand how the physics and 
evolution of baryons within the halos can explain the properties of 
high-$z$ ETGs, and how it is possible to build-up massive ETGs at
higher redshifts than expected.

The models incorporating a new treatment of the feedback 
processes (including AGN), or the ones based on hydrodynamic simulations, 
seem to obtain a better agreement with the observations \cite{gra,menci,nag,
silva}. The most recent simulations \cite{springel} are also capable to 
predict a downsizing evolutionary pattern for the star formation, but
most of the mass assembly still occurs at moderately low redshifts
($0<z<1$), and a specific comparison with the observed properties and
evolution of the ETGs has not been published yet \cite{delucia,bower}. 
It is interesting to mention that very recent hydrodynamic simulations
showed that it is possible to build a realistic massive
(${\cal M} \sim 10^{11}$ M$_{\odot}$) giant elliptical galaxy with a plausible 
formation history without requiring neither recent major mergers nor feedback
from supernovae or AGN, but simply starting from appropriate $\Lambda$CDM 
initial conditions \cite{naab}. 

\section{Searching for the progenitors}

The spectral and color properties of the distant, old ETGs discussed 
in Section 3 imply unambiguously a star formation history with: strong 
($>100$ M$_{\odot}$ 
yr$^{-1}$) and short-lived ($\tau \sim$0.1-0.3 Gyr) starburst 
(where SFR$\propto exp(t/ \tau)$), the onset of star formation 
occurring at high redshift ($z_{f}>2-3$), and a passive--like evolution 
after the major starburst \cite{cim04,da05a,mcc,long05}. In addition,
the ETG precursors should also have the strong clustering expected
in the $\Lambda$CDM models for the populations located in massive
dark matter halos and strongly biased environments. Recent observations 
have indeed uncovered galaxies matching the above requirements.

\subsection{$BzK$--selected galaxies}

The K20 survey unveiled a population of star-forming galaxies 
at $z\sim2$ which have strong star formation rates of $\sim 
200-300$ M$_{\odot}$ yr$^{-1}$, substantial dust extinction with 
$E(B-V)\sim 0.3-0.6$, stellar masses up to $10^{11}$ M$_{\odot}$, 
merging-like rest-frame UV morphology becoming more compact in 
the rest-frame optical, strong clustering with $r_0 \sim$8--12 Mpc
depending on the adopted redshift distribution, and possibly high 
metallicity \cite{da04a,da04b,da05b,de04,kong}. These starburst 
galaxies can be efficiently selected at $1.4<z<2.5$ with a ``reddening-free'' 
color--color diagram (z-$K_s$ vs. $B-$z; the $B$z$K$ selection) \cite{da04b}. 
The $B$z$K$--selected starbursts with $K_s<20$ are on 
average more reddened, massive and star-forming than the optically-selected 
star-forming galaxies, but they show a significant overlap with other 
star-forming galaxy populations in the same redshift range selected based 
on other optical/near-IR criteria \cite{reddy} or on mid--infrared - to - 
radio selections \cite{da05b,dann}. The properties of the 
$B$z$K$--selected starbursts suggest that these galaxies are massive 
galaxies caught during their major activity of mass assembly and 
star-formation, possibly being the progenitors of the present-day massive 
and metal-rich ETGs. At the stellar mass threshold of ${\cal M}>10^{11} 
M_{\odot}$, the number density of these $B$z$K$--selected starbursts is 
comparable with that of old passively evolving galaxies in the same 
redshift range of $1.4<z<2.5$ \cite{da05b,mcc,kong}. 

\subsection{Other precursor candidates}

Other massive galaxy progenitor candidates at $z>2$ characterized by
old stellar populations and/or strong star formation, large stellar 
and/or dynamical masses and high metallicity include the submm/mm-selected 
galaxies \cite{cha,danner,swin,gre}, the ``Distant Red Galaxies'' selected with 
$J-K_s>2.3$ \cite{franx,vd04,pap}, the optically-selected 
BM/BX/LBG systems with 
bright $K$-band fluxes \cite{ade,reddy,sha}, the IRAC Extremely Red Objects
(IEROs) \cite{yan}, the HyperEROs \cite{totani,sar04}, and a fraction of 
dusty EROs \cite{yan05}.

\subsection{The GMASS project} 

GMASS ({\it ``Galaxy Mass Assembly ultra-deep Spectroscopic
Survey''}) is a new ongoing spectroscopic project based on an ESO 
Large Program (PI A. Cimatti) aimed at doing ultradeep spectroscopy with 
VLT+FORS2 of galaxies selected at $4.5\mu$m with Spitzer+IRAC adopting 
two simple criteria: $m(4.5\mu$m)$<23.0$ (AB) and $z_{phot}>1.4$.
The integration times are very deep (15-30 hours per spectroscopic mask). 
The main scientific driver of {\it GMASS} is the spectroscopic
identification and study of the high-$z$ progenitors of the massive ETGs 
in a way complementary to other surveys and with a selection more
sensitive to the rest-frame near-IR emission (i.e. stellar mass)
up to $z\sim3$. 

\begin{acknowledgements}
I am grateful to Emanuele Daddi for his comments and to the K20 and 
GMASS collaborators for their invaluable contributions to the projects.
\end{acknowledgements}

\bibliographystyle{aa}

\end{document}